\documentclass[cits]{PoS}

\def\Dsl{{\rlap{\kern2.25pt /}{D}}}
\def\Asl{{\rlap{\kern2.25pt /}{A}}}
\def\dsl{{\rlap{\kern0.5pt /}{\partial}}}
\def\xisl{{\rlap{\kern0.5pt /}{\xi}}}
\def\asl{{\rlap{\kern0.5pt /}{a}}}
\def\bsl{{\rlap{\kern0.5pt /}{b}}}

\def\Tr{{\rm Tr}}
\def\({\left(}
\def\){\right)}

\usepackage{wrapfig}
\usepackage{amssymb, amsthm, amsmath}

\title{Exotic phases of finite temperature SU(N) gauge theories with massive fermions: F, Adj, A/S }

\ShortTitle{Exotic phases of finite temperature SU(N) gauge theories with massive fermions: F, Adj, A/S }

\author{\speaker{Joyce Myers} and Michael Ogilvie%
\\%         \thanks{A footnote may follow.}\\
        Washington University in St. Louis\\
        E-mail: \email{jcmyers@wustl.edu},
        \email{mco@wuphys.wustl.edu}}

\abstract{The phase diagrams at high temperature of $SU(N)$ gauge theories with massive fermions are calculated by numerically minimizing the one-loop effective potential. We consider fermions in the Fundamental (F), Adjoint (Adj), Antisymmetric (AS), and Symmetric (S) representations, for N from 3 to 9, with periodic and antiperiodic boundary conditions applied. For one flavour of AS/S (Dirac) fermion with periodic boundary conditions the C-breaking phase is favoured perturbatively for all values of the fermion mass. In the case of one flavour of adjoint Majorana fermion, and periodic boundary conditons, the deconfined phase is favoured for any fermion mass. For one or more adjoint Dirac fermion (two or more Majorana fermions) we find partially-confining phases as well as new phases with unusual properties. Our results for $SU(3)$ and $SU(4)$ are consistent with our lattice simulations of a related model.}

\FullConference{The XXVI International Symposium on Lattice Field Theory \\
		 July 14 - 19, 2008\\
		 Williamsburg, Virginia, USA}

\begin{document}

\begin{section}{Introduction}
%%%%%%%%%%%%%%%%%%%%%%%%%%%%%%%%%%%%%%%%%%%%%%
Certain QCD-like theories can be useful for studying confinement and chiral symmetry breaking. One of these is adjoint QCD [QCD(Adj)] \cite{Davies:1999uw,Unsal:2007jx} which is $SU(N)$ gauge theory with adjoint representation fermions\footnote{In this paper we further define adjoint QCD with periodic boundary conditions (PBC,+) on fermions rather than the usual antiperiodic boundary conditions (ABC,-). This results in additional phases and allows for comparison of lattice results with analytic calculations.} instead of fundamental. For a theory which approximates adjoint QCD our lattice results in \cite{Myers:2007vc} indicated that the confined phase could be accessed perturbatively\footnote{For a related theory the lattice results in \cite{Dumitru:2007ir} also suggest accessibility of the confined phase above $T_{c,gauge}$.}. It was shown that this result is well supported by perturbation theory in the high temperature regime. In this paper we show that for $SU(N)$ gauge theories with fermions in various representations varying the fermion mass gives rise to non-trivial phase structure when considering PBC on fermions.

In our earlier work \cite{Myers:2007vc} simulations were performed using an extension to Yang-Mills theory of an adjoint Polyakov loop term, which is like adding a heavy adjoint quark. Using one-loop perturbation theory we consider two new extensions to Yang-Mills theory: 1) multiply wound adjoint Polyakov loops (center-stabilized Yang-Mills theory), 2) fermions in the adjoint representation with nonzero mass.

We considered various $N$ and $N_f$ as well as other representations of fermions: fundamental (F), antisymmetric (AS), and symmetric (S), and refer the reader to \cite{Myers:2008jm} for details of all our results.
%%%%%%%%%%%%%%%%%%%%%%%%%%%%%%%%%%%%%%%%%%%%%%%
\end{section}
\begin{section}{Center-stabilized Yang-Mills theory}

Perturbative accessibility of the confined phase for all $N$ is possible for certain types of $Z(N)$-invariant extensions to Yang-Mills theory. In \cite{Ogilvie:2007tj} we introduced an extension in terms of powers of the Polyakov loop $P = {\rm diag} \{ e^{i v_1}, e^{i v_2}, ... , e^{i v_N} \}$, 

\begin{equation}
V_{ext} (P) \equiv \frac{1}{\beta} \sum_{n=1}^{\lfloor N/2 \rfloor} a_n \, \Tr_F \left( P^n \right) \Tr_F ( P^{\dagger n} ) = \frac{1}{\beta} \sum_{n=1}^{\lfloor N/2 \rfloor} a_n \sum_{i, \, j=1}^{N} \cos \left[ n \left( v_i - v_j \right) \right]
\end{equation}

\noindent where $\lfloor N/2 \rfloor$ is the integer part of $N/2$. This is the minimum number of terms required to obtain the confined phase for some value of the $a_n$ parameters.

Including the boson contribution \cite{Gross:1980br} from pure Yang-Mills theory

\begin{equation}
V_{CYM} (P) =  - \frac{2}{\pi^2 \beta^4} \sum_{n=1}^{\infty} \frac{1}{n^4} \left[ \Tr_A \( P^n \) \right] + \frac{1}{\beta} \sum_{n=1}^{\lfloor N/2 \rfloor} a_n \Tr_F (P^n) \Tr_F (P^{\dagger n}) .
\label{cym_eqn}
\end{equation}

\noindent This potential has recently been studied more extensively in \cite{Unsal:2008ch} and we have therefore adopted their notation ("$a_n$") and nomenclature ("center-stabilized Yang-Mills theory") in this paper. We minimized $V_{CYM}$ with respect to the Polaykov loop eigenvalues $v_i$ to determine the phase diagram for a range of values of the $a_n$.

\end{section}
%%%%%%%%%%%%%%%%%%%%%%%%%%%%%%%%%%%%%%%%%%%
\begin{section}{One-loop effective potential with massive fermions in rep $R$}

The one-loop effective potential for $N_f$ Majorana fermions ($N_{f, Dirac} = \frac{1}{2} N_{f}$) of mass $m$ and in representation $R$ in a background Polyakov loop $P$ is \cite{Meisinger:2001fi}:

\begin{equation}
\begin{aligned}
V_{1-loop} (P, m) \equiv &- \frac{1}{\beta V_3} \ln Z(P, m)\\
= &\frac{1}{\beta V_3} \left[ - N_f \ln \det \left( -D_R^2 (P) + m^2 \right) + \ln \det \left(-D_{adj}^2 (P) \right) \right]\\
= & \frac{m^2 N_f}{\pi^2 \beta^2} \sum_{n=1}^{\infty} \frac{(\pm 1)^n}{n^2} {\mathrm Re} \left[ \Tr_R \left( P^n \right) \right] K_2 \left( n \beta m \right) - \frac{2}{\pi^2 \beta^4} \sum_{n=1}^{\infty} \frac{1}{n^4} \Tr_A \( P^n \)
\end{aligned}
\label{v_1loop}
\end{equation}

\noindent where we have $(+ 1)^n$ for PBC and $(- 1)^n$ for ABC applied to fermions. To obtain the preferred phases for a range of $m \beta$ we numerically minimize $V_{1-loop}$ with respect to the eigenvalue angles $v_i$ of the Polyakov loop.

\end{section}
%%%%%%%%%%%%%%%%%%%%%%%%%%%%%%%%%%%%%%%%%%%
\begin{section}{Results}
%%%%%%%%%%%%%%%%%%%%%%%%%%%%%%%%%%%%%%%%%%%
\begin{subsection}{Phases of adjoint QCD, PBC on fermions , $N_f > 1$ Majorana flavour}

\begin{figure}
  \hfill
    \begin{minipage}[t]{.45\textwidth}
    \begin{center}
\includegraphics[width=0.8\textwidth]{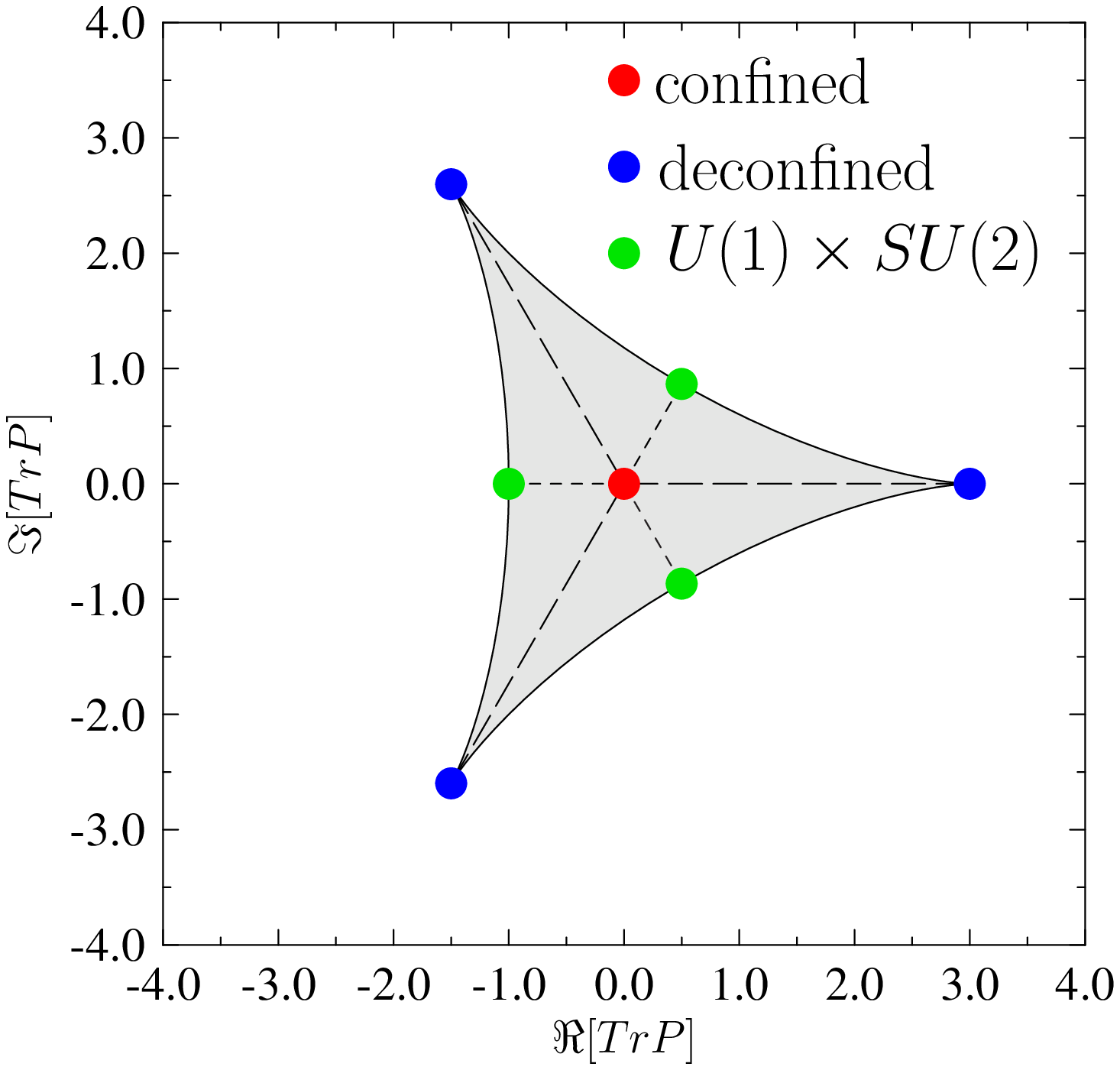}
       \caption{$N_c = 3$ phases of QCD(Adj)}
       \label{adj_phases_n3}
    \end{center}
  \end{minipage}
  \hfill
  \begin{minipage}[t]{.45\textwidth}
    \begin{center}
\includegraphics[width=0.8\textwidth]{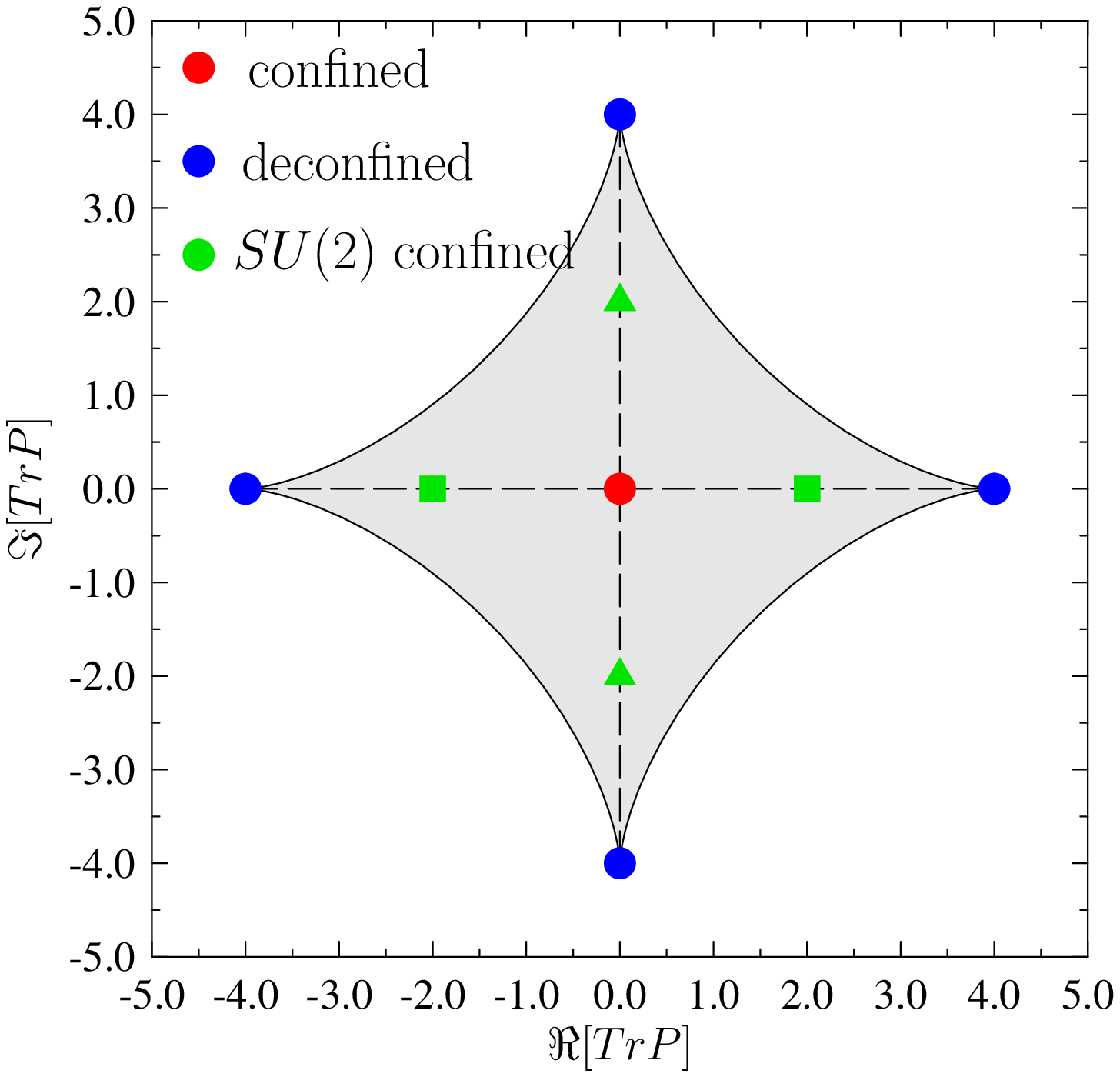}
\caption{$N_c = 4$ phases of QCD(Adj)}
\label{adj_phases_n4}
    \end{center}
  \end{minipage}
  \hfill
\end{figure}

The phase diagram of $SU(N)$ gauge theories with adjoint fermions is quite rich when periodic boundary conditions are applied to at least two Majorana fermion flavours. Figures \ref{adj_phases_n3} - \ref{adj_phases_n6} show the observed phases for $N = 3 - 6$. In all cases the confined phase is observed. The confined phase is conventiently defined in terms of the $N$ Polyakov loop eigenvalue angles $v_i$,

\begin{equation}
\begin{aligned}
\text{confined:} \hspace{5mm} &{\bf v} = \{ 0, \frac{2 \pi}{N}, \frac{4 \pi}{N}, ... , \frac{2 \pi (N-1)}{N} \} \hspace{2mm} N \,\, \text{odd}\\
&{\bf v} = \{ \frac{\pi}{N}, \frac{3 \pi}{N}, ... , \frac{ (2 N - 1) \pi}{N} \} \hspace{6mm} N \,\, \text{even} .
\end{aligned}
\label{confinedphase}
\end{equation}

\noindent The deconfined phases are also observed:

\begin{equation}
\text{deconfined:} \hspace{3mm} {\bf v} = \{ 0, 0, ... , 0 \} \,\, \text{and all $N - 1$ nontrivial $Z(N)$ rotations} .
\end{equation}

In $SU(3)$ there are additional $SU(2) \times U(1)$ (or "skewed") phases:

\begin{equation}
SU(2) \times U(1): {\bf v} = \{ 0, \pi, \pi \} \,\, \text{and $Z(3)$ rotations} .
\end{equation}

For $SU(4)$ the new phase is a partially-confined $Z(2)$-invariant phase:

\begin{equation}
SU(2) \,\, \text{conf:} \,\, {\bf v} = \{ 0, 0, \pi, \pi \} \,\, \text{and} \,\, \{ \frac{\pi}{2}, \frac{\pi}{2}, \frac{3 \pi}{2}, \frac{3 \pi}{2} \} .
\end{equation}

\noindent In Figures \ref{adj_phases_n4} and \ref{adj_phases_n6} the use of a shape other than a circle indicates that the vacua occur together only. For example, in Figure \ref{adj_phases_n4} for the $SU(2)$ confined phase the two vacua on the real axis (represented by squares) only occur together such that $\Tr_F P = 0$, however, $\Tr_F P^2 \ne 0$. The same is true of the vacua on the imaginary axis (represented by triangles).

\begin{figure}
\begin{minipage}[t]{0.45\textwidth}
\begin{center}
\includegraphics[width=0.8\textwidth]{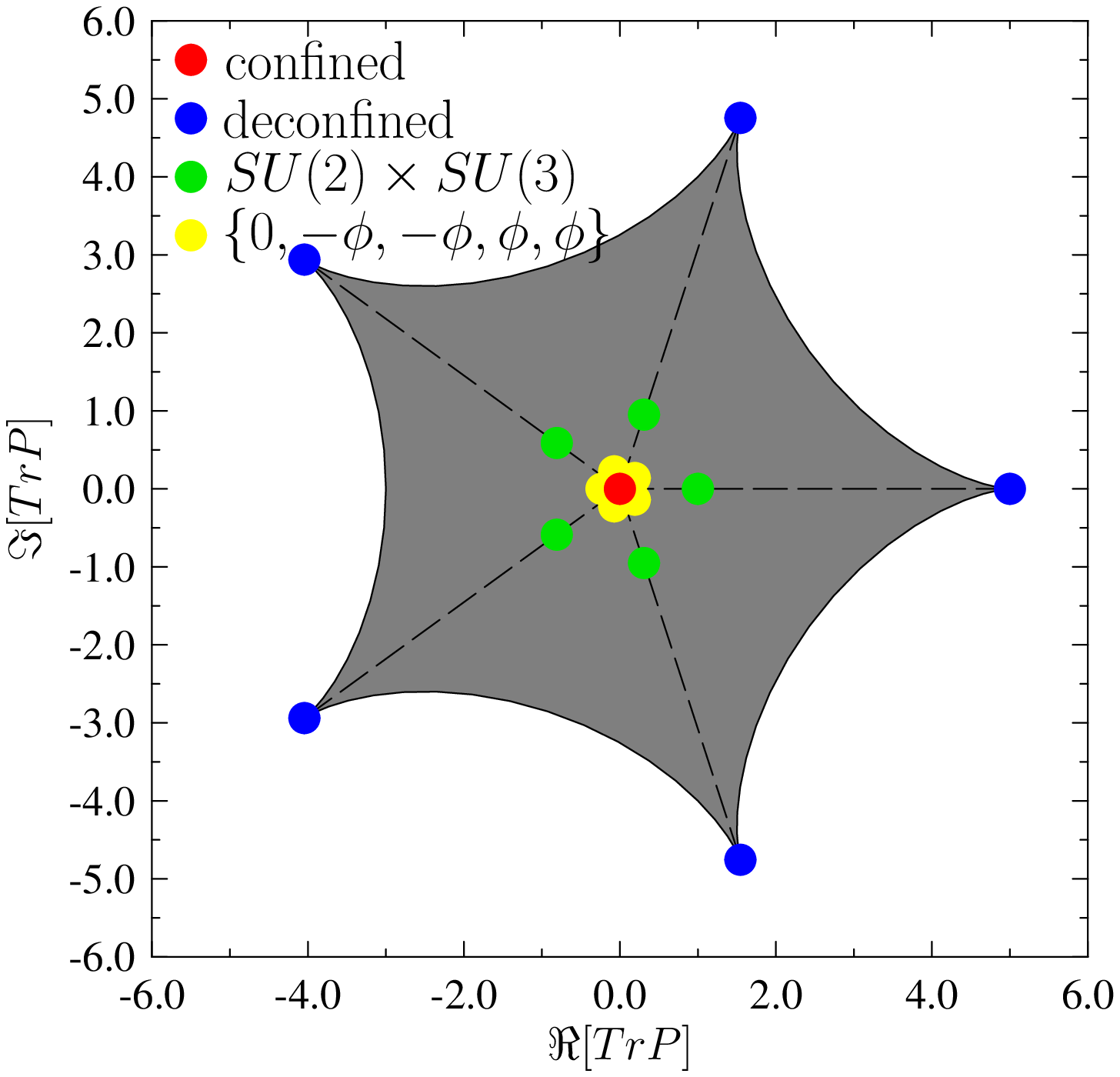}
\caption{$N = 5$ phases of QCD(Adj)}
\label{adj_phases_n5}
\end{center}
\end{minipage}
\hfill
\begin{minipage}[t]{0.45\textwidth}
\begin{center}
\includegraphics[width=0.8\textwidth]{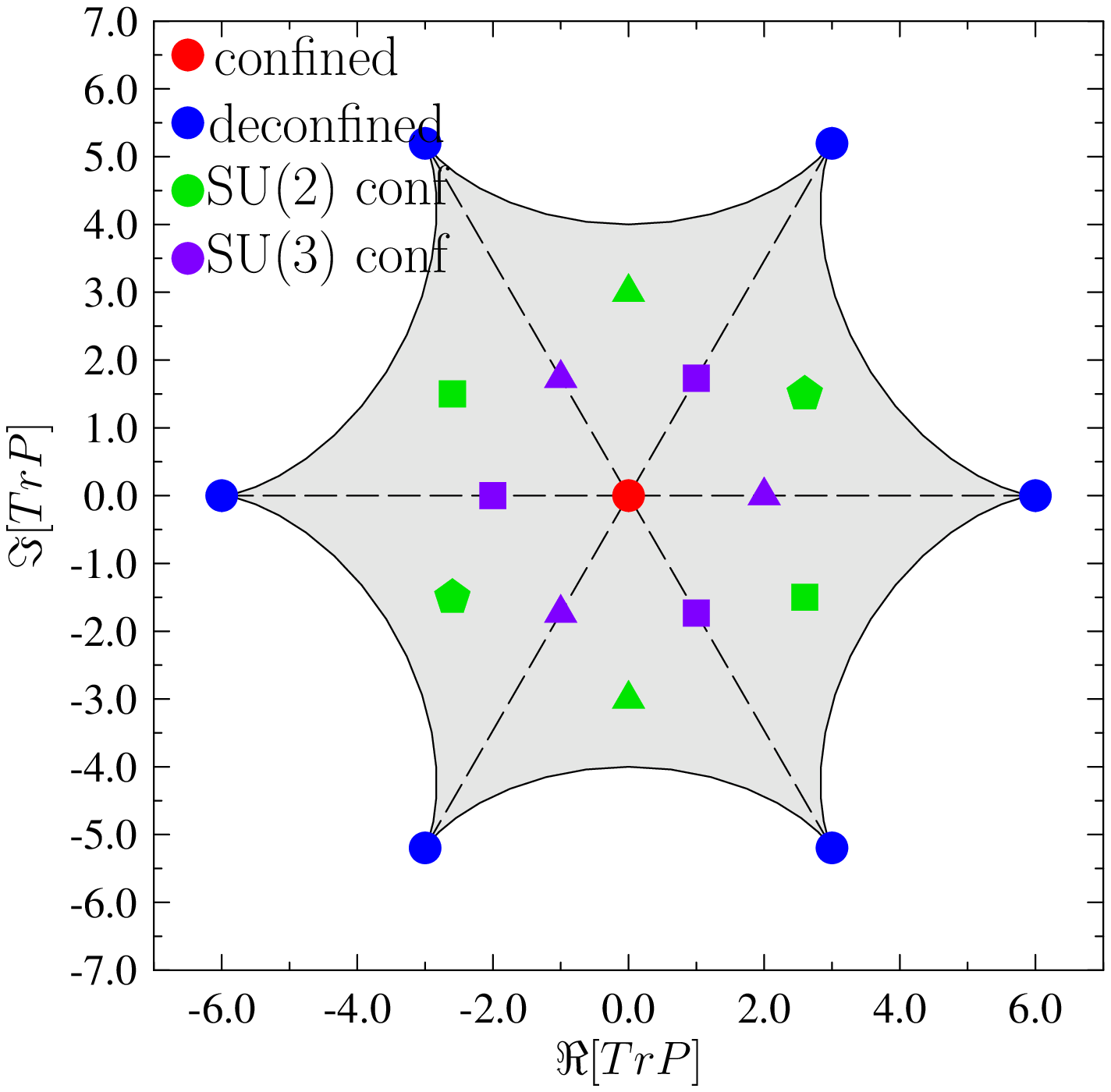}
\caption{$N = 6$ phases of QCD(Adj)}
\label{adj_phases_n6}
\end{center}
\end{minipage}
\hfill
\end{figure}

When $N = 5$ the new phases are the $SU(2) \times SU(3)$ phase, and an $SU(2) \times SU(2) \times U(1)$ phase:

\begin{equation}
\begin{aligned}
&SU(2) \times SU(3): \,\, {\bf v} = \{ 0, 0, 0, \pi, \pi \} \,\, \text{and $Z(5)$ rotations}\\
&SU(2) \times SU(2) \times U(1): \,\, {\bf v} = \{ 0, - \phi, - \phi, \phi, \phi \} \,\, \text{and $Z(5)$ rotations} .
\end{aligned}
\end{equation}

\noindent The $SU(2) \times SU(2) \times U(1)$ phase is unique in that the Polyakov loop eigenvalues are not constant as $m \beta$ as varied, but rather $\phi$ decreases as $m \beta$ increases causing the Polyakov loop eigenvalues to be attracted together.

When $N = 6$ the new phases are both partially confined: the $SU(2)$-confined phase, and an $SU(3)$-confined phase:

\begin{equation}
\begin{aligned}
&SU(2) \,\, \text{conf:} \,\, {\bf v} = \{ \frac{\pi}{2}, \frac{3 \pi}{2}, \frac{\pi}{2}, \frac{3 \pi}{2}, \frac{\pi}{2}, \frac{3 \pi}{2} \}, \{ \frac{5 \pi}{6}, \frac{11 \pi}{6}, \frac{5 \pi}{6}, \frac{11 \pi}{6}, \frac{5 \pi}{6}, \frac{11 \pi}{6} \}, \{ \frac{7 \pi}{6}, \frac{\pi}{6}, \frac{7 \pi}{6}, \frac{\pi}{6}, \frac{7 \pi}{6}, \frac{\pi}{6} \}\\
&SU(3) \,\, \text{conf:} \,\, {\bf v} = \{ 0, \frac{2 \pi}{3}, \frac{4 \pi}{3}, 0, \frac{2 \pi}{3}, \frac{4 \pi}{3} \}, \{ \frac{\pi}{3}, \pi, \frac{5 \pi}{3}, \frac{\pi}{3}, \pi, \frac{5 \pi}{3} \} .
\end{aligned}
\label{su6newphases}
\end{equation}

\end{subsection}
%%%%%%%%%%%%%%%%%%%%%%%%%%%%%%%%%%%%%%%%%%%
\begin{subsection}{Results for Adjoint QCD, PBC on fermions, $N_f = 2$ Majorana flavours}

Figures \ref{v_adj_1loop_n3}(L), \ref{v_adj_1loop_n4}(L), \ref{v_adj_1loop_n5}(L), and \ref{adj_n6}(L) show the phase diagram of adjoint QCD with PBC on fermions and $N_f = 2$ Majorana flavours as a function of $m \beta$ for $N = 3 - 6$. The black dots indicate the result of numerical minimization of $V_{1-loop}$ in eq. (\ref{v_1loop}) with respect to the eigenvalue angles $v_i$. The coloured lines result from plugging into eq. (\ref{v_1loop}) the known eigenvalue angles from eqs. (\ref{confinedphase} - \ref{su6newphases}). The coloured line on which the dots lie tells us the preferred phase for a value of $m \beta$.

\begin{figure}
  \hfill
  \begin{minipage}[t]{.45\textwidth}
    \begin{center}  
      \includegraphics[width=0.9\textwidth]{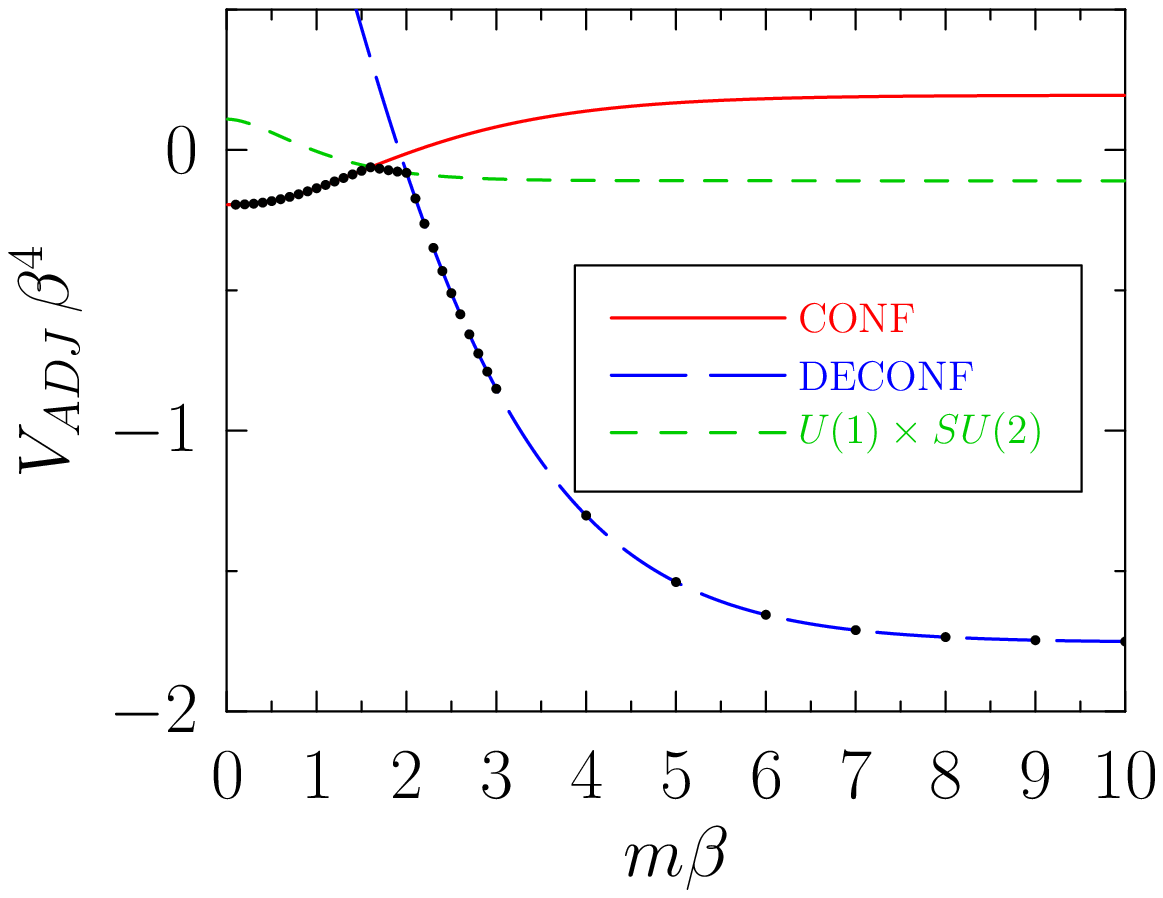}
%      \caption{$V_{ADJ(+)}$, $N = 3$, $N_f = 2$ Majorana flavours}
%      \label{v_adj_1loop_n3}
    \end{center}
  \end{minipage}
  \hfill
  \begin{minipage}[t]{.45\textwidth}
    \begin{center}
\includegraphics[width=0.9\textwidth]{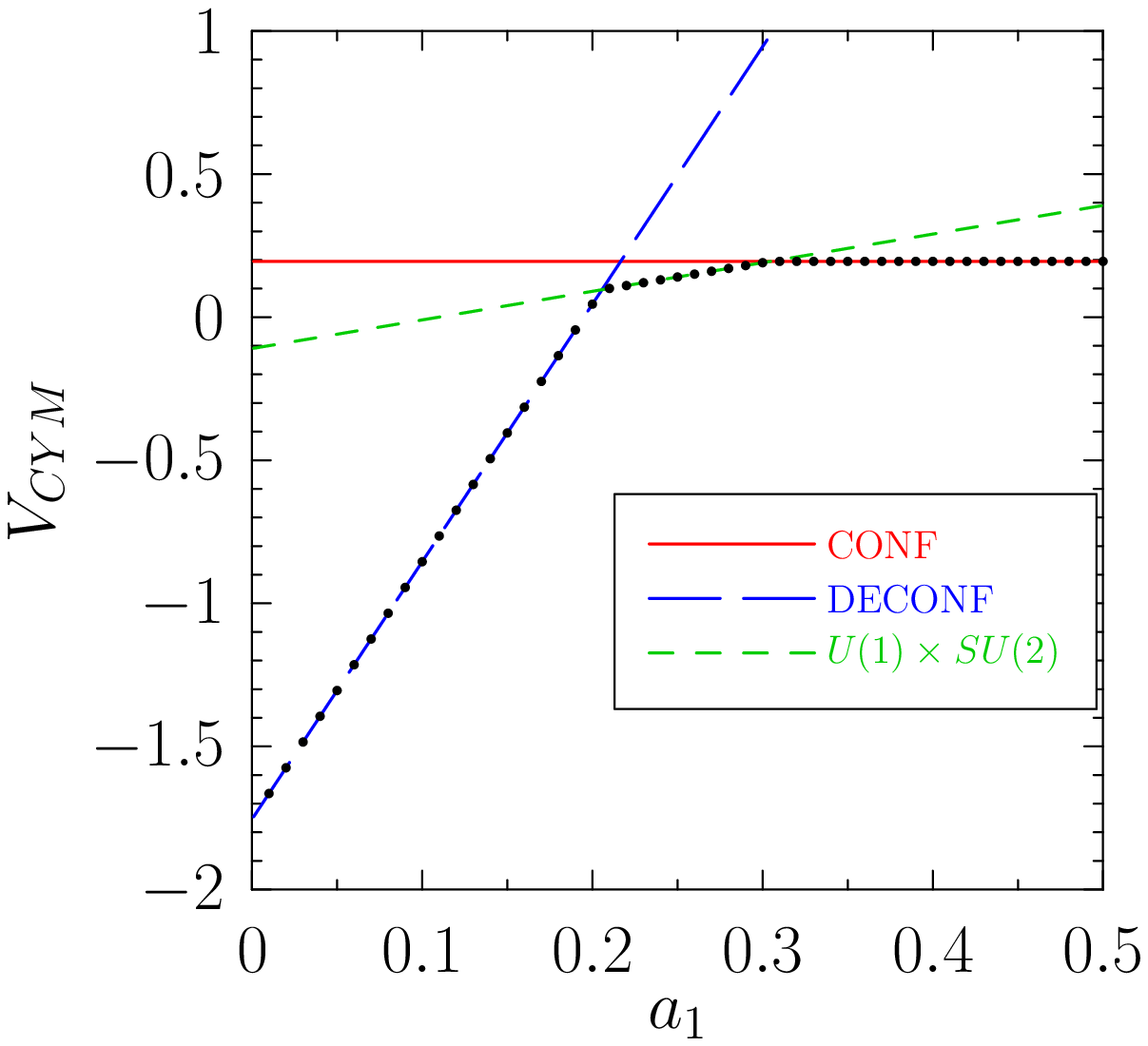}
%      \caption{$V_{CYM}$ vs. $a_1$, $N = 3$}
%      \label{v_cym_n3}
    \end{center}
  \end{minipage}
  \hfill
  \caption{$N = 3$: (Left) $V_{ADJ(+)}$, $N_f = 2$ Majorana flavours; (Right) $V_{CYM}$ vs. $a_1$}
  \label{v_adj_1loop_n3}
\end{figure}

\begin{figure}
  \hfill
  \begin{minipage}[t]{.45\textwidth}
    \begin{center}  
      \includegraphics[width=0.9\textwidth]{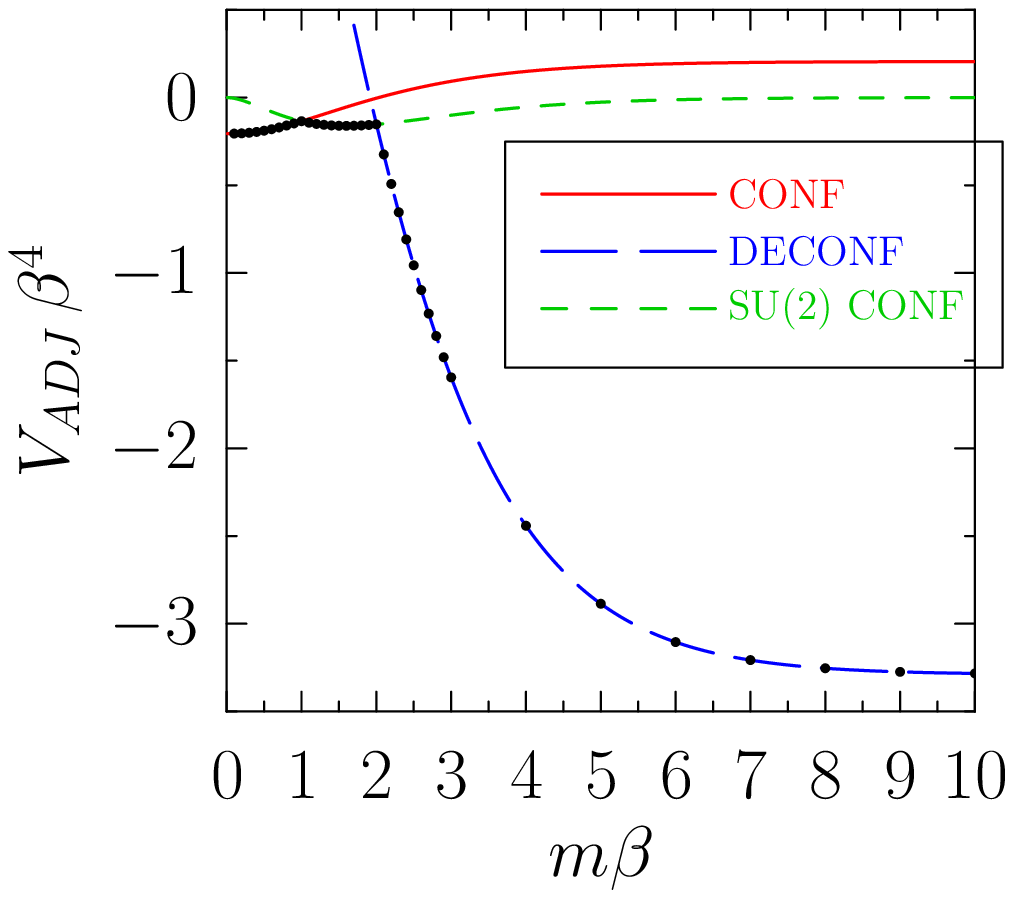}
%      \caption{$V_{ADJ(+)}$, $N = 4$, $N_f = 2$ Majorana flavours}
%      \label{v_adj_1loop_n4}
    \end{center}
  \end{minipage}
  \hfill
  \begin{minipage}[t]{.45\textwidth}
    \begin{center}
\includegraphics[width=0.9\textwidth]{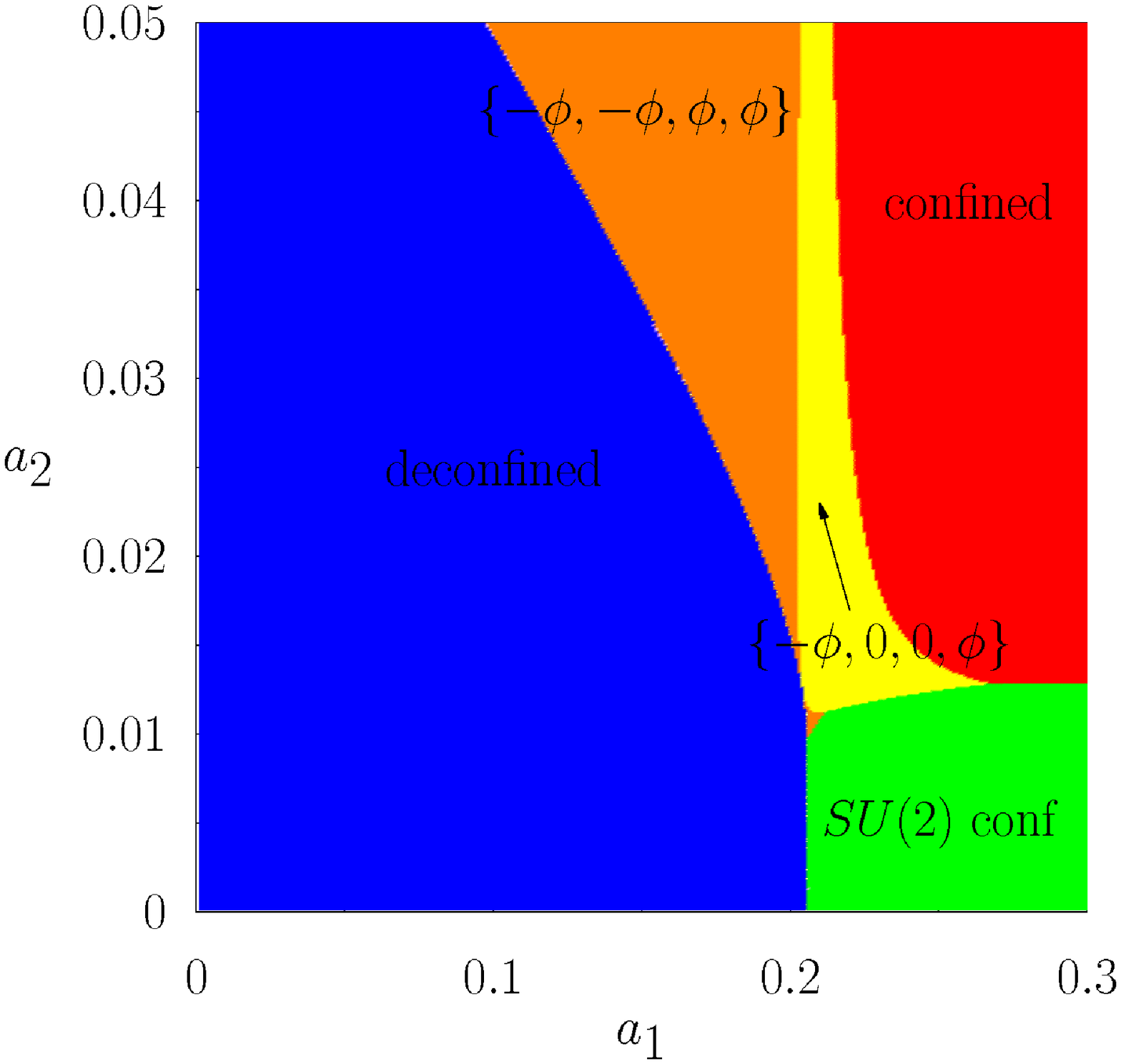}
%\includegraphics[width=0.9\textwidth]{mod_SU4.eps}
%      \caption{phase diagram of $V_{CYM}$ in the $a_1 - a_2$ plane, $N = 4$}
%      \label{v_cym_n4}
    \end{center}
  \end{minipage}
  \hfill
\caption{$N = 4$: (Left) $V_{ADJ(+)}$, $N_f = 2$ Majorana flavours; (Right) phase diagram of $V_{CYM}$ in the $a_1 - a_2$ plane}
\label{v_adj_1loop_n4}
\end{figure}

\begin{figure}
  \hfill
  \begin{minipage}[t]{.45\textwidth}
    \begin{center}  
      \includegraphics[width=0.9\textwidth]{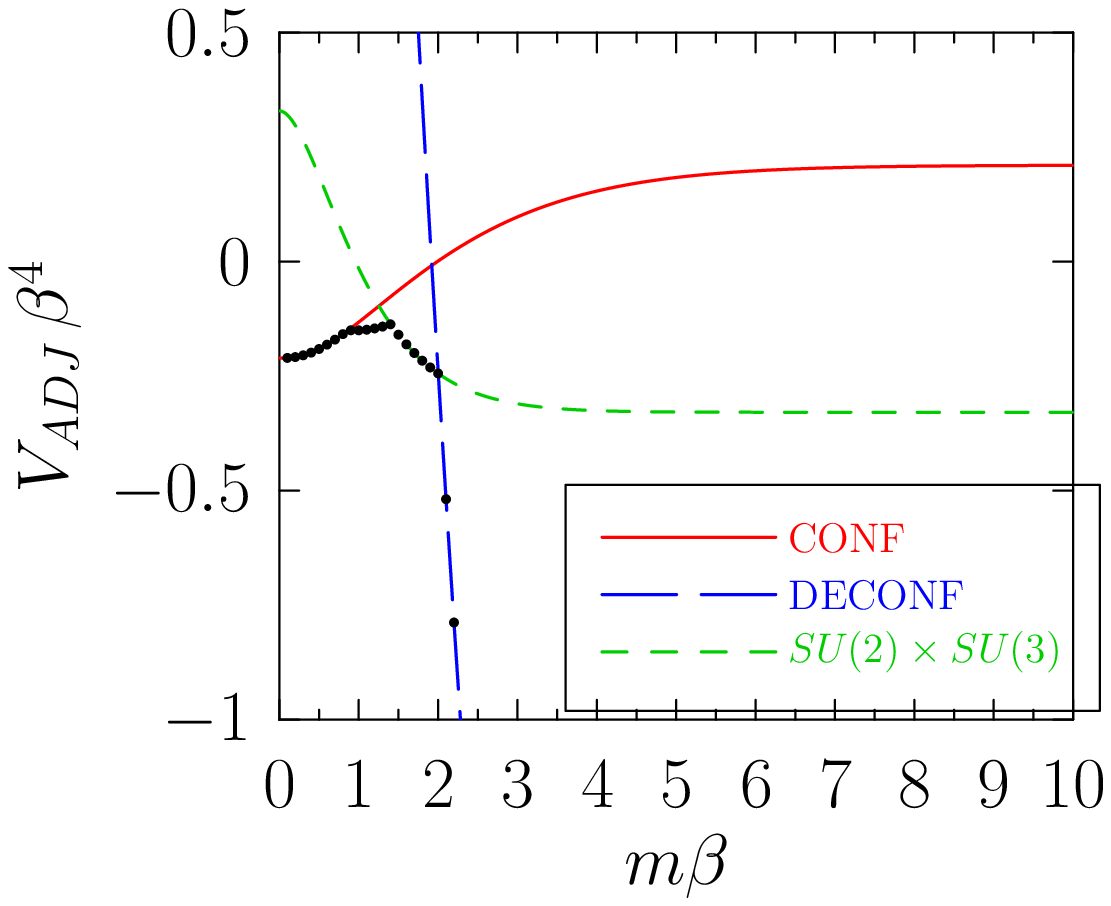}
%      \caption{$V_{ADJ(+)}$, $N = 5$, $N_f = 2$ Majorana flavours}
%      \label{v_adj_1loop_n5}
    \end{center}
  \end{minipage}
  \hfill
  \begin{minipage}[t]{.45\textwidth}
    \begin{center}
\includegraphics[width=0.9\textwidth]{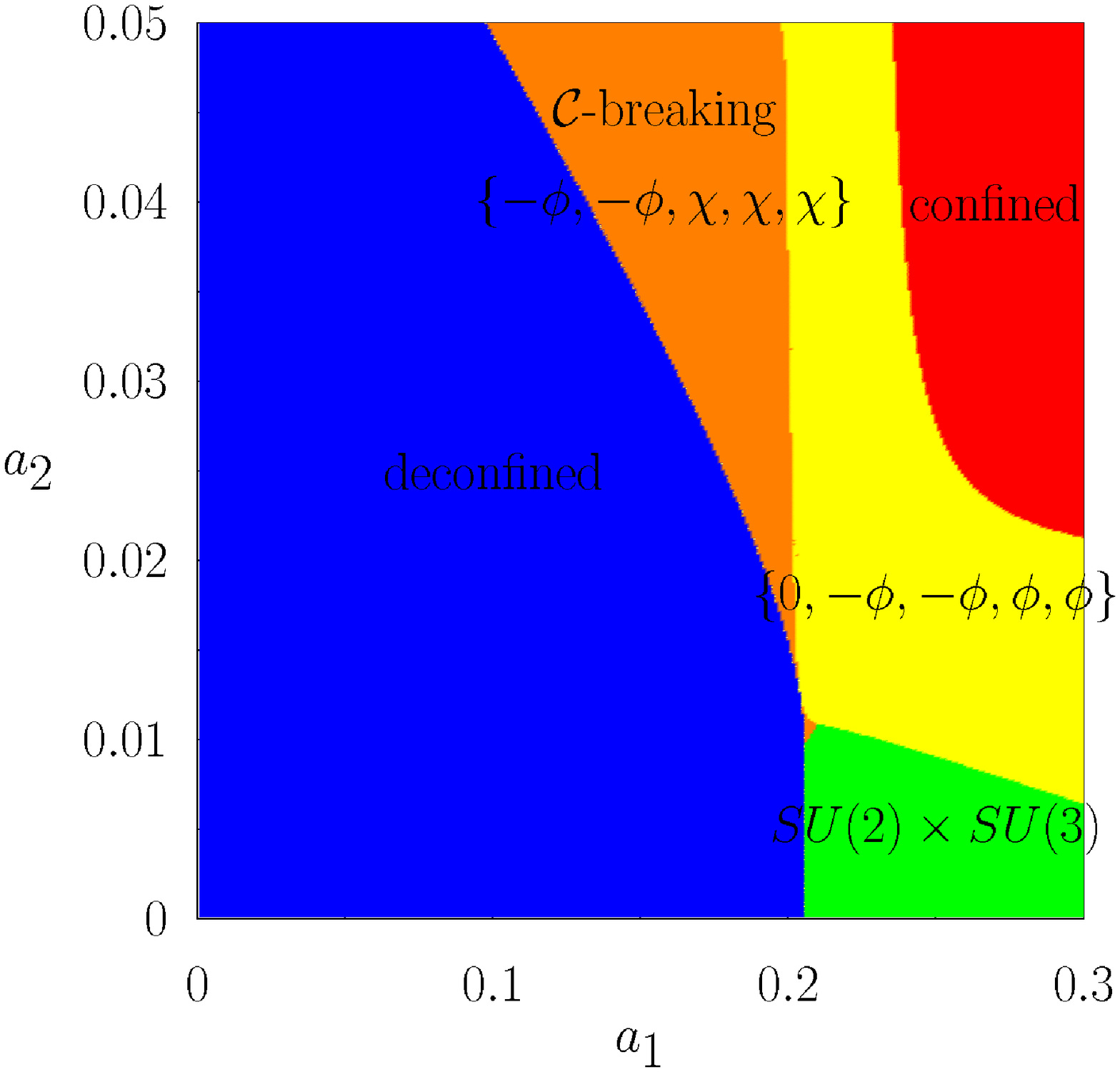}
%\includegraphics[width=0.9\textwidth]{mod_SU5.eps}
%      \caption{phase diagram of $V_{CYM}$ in the $a_1 - a_2$ plane, $N = 5$}
%      \label{v_cym_n5}
    \end{center}
  \end{minipage}
  \hfill
\caption{$N = 5$: (Left) $V_{ADJ(+)}$, $N_f = 2$ Majorana flavours; (Right) phase diagram of $V_{CYM}$ in the $a_1 - a_2$ plane}
\label{v_adj_1loop_n5}
\end{figure}

In the case of $N = 5$, Figure \ref{v_adj_1loop_n5}(L) shows that the $SU(2) \times SU(2) \times U(1)$ phase does not have a corresponding theory curve because the Polyakov loop eigenvalue angles ${\bf v} = \{ 0, - \phi, - \phi, \phi, \phi \}$ change with with $m \beta$.

As $m \beta$ is increased, Figures \ref{v_adj_1loop_n3}(L), \ref{v_adj_1loop_n4}(L), \ref{v_adj_1loop_n5}(L), and \ref{adj_n6}(L) indicate that the phases in Figures \ref{adj_phases_n3} - \ref{adj_phases_n6} are traversed in order of increasing $\left| \Tr_F P \right|$ (considering only one of the vacua in the case of the partially confined phases). With each transition the Polyakov loop eigenvalues are increasingly repelled going from the confined phase, through the new phases, to the deconfined phase.

Another important observation is that the confined phase is less accessible perturbatively as $N$ increases. For $N = 3$, Figure \ref{v_adj_1loop_n3}(L) indicates that the confined phase is accessible for $m \beta \le 1.6$. For $N = 6$, Figure \ref{adj_n6}(L) shows that it is only accessible for $m \beta \le 0.6$. Considering the large $N$ limit Figure \ref{conf} shows $(m \beta)_{crit}$, the maximum value of $m \beta$ (with uncertainty $+0.1$) for which the confined phase is accessible, for $N$ from $2$ to $18$. It indicates that the range of $m \beta$ for which the confined phase is accessible shrinks rapidly at first, then levels off for large $N$.

The decrease in the range of $m \beta$ for which the confined phase is perturbatively accessible as $N \rightarrow \infty$ can be partially offset by increasing the number of fermion flavours $N_f$. Figures \ref{adj_n6}(L) and \ref{adj_n6}(R) give the case of $N = 6$; as $N_f$ is increased from $2$ to $3$, the range of $m \beta$ for which the confined phase is accessible increases as well. However, there is a natural limit in that $N_f \le 5$ Majorana fermion flavours are required to preserve asymptotic freedom.

Figures \ref{v_adj_1loop_n3}(R), \ref{v_adj_1loop_n4}(R), and \ref{v_adj_1loop_n5}(R) show the results of minimizing $V_{CYM}$ of eq. (\ref{cym_eqn}) with respect to $a_1$ for $N = 3$, and with respect to $a_1$ and $a_2$ in the case of $N = 4$ and $5$. As shown from side-by-side comparison with Figures \ref{v_adj_1loop_n3}(L), \ref{v_adj_1loop_n4}(L), and \ref{v_adj_1loop_n5}(L), the center-stabilized model always includes the phases QCD(Adj), as well as additional phases in the case of $N = 4$ and $5$. But, the additional phases can always be circumnavigated by choosing an appropriate path through the $a_n$ space, allowing traversal of the phases in the same order as they appear in QCD(Adj) for increasing $m \beta$.

\begin{figure}
  \hfill
  \begin{minipage}[t]{.45\textwidth}
    \begin{center}  
      \includegraphics[width=0.9\textwidth]{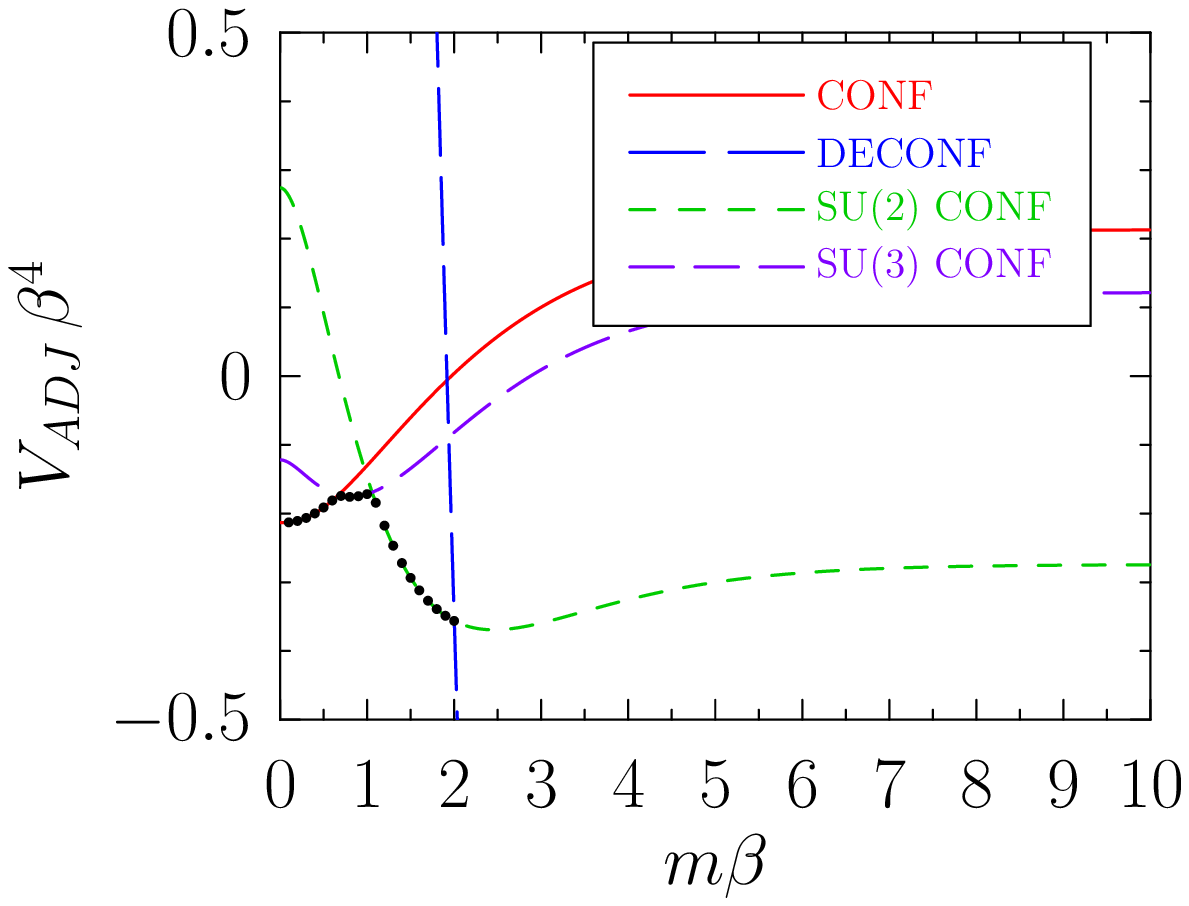}
%      \caption{$V_{ADJ(+)}$ for $N = 6$, $N_f = 2$ Majorana flavours}
%      \label{adj_n6_nf2}
    \end{center}
  \end{minipage}
  \hfill
  \begin{minipage}[t]{.45\textwidth}
    \begin{center}
\includegraphics[width=0.9\textwidth]{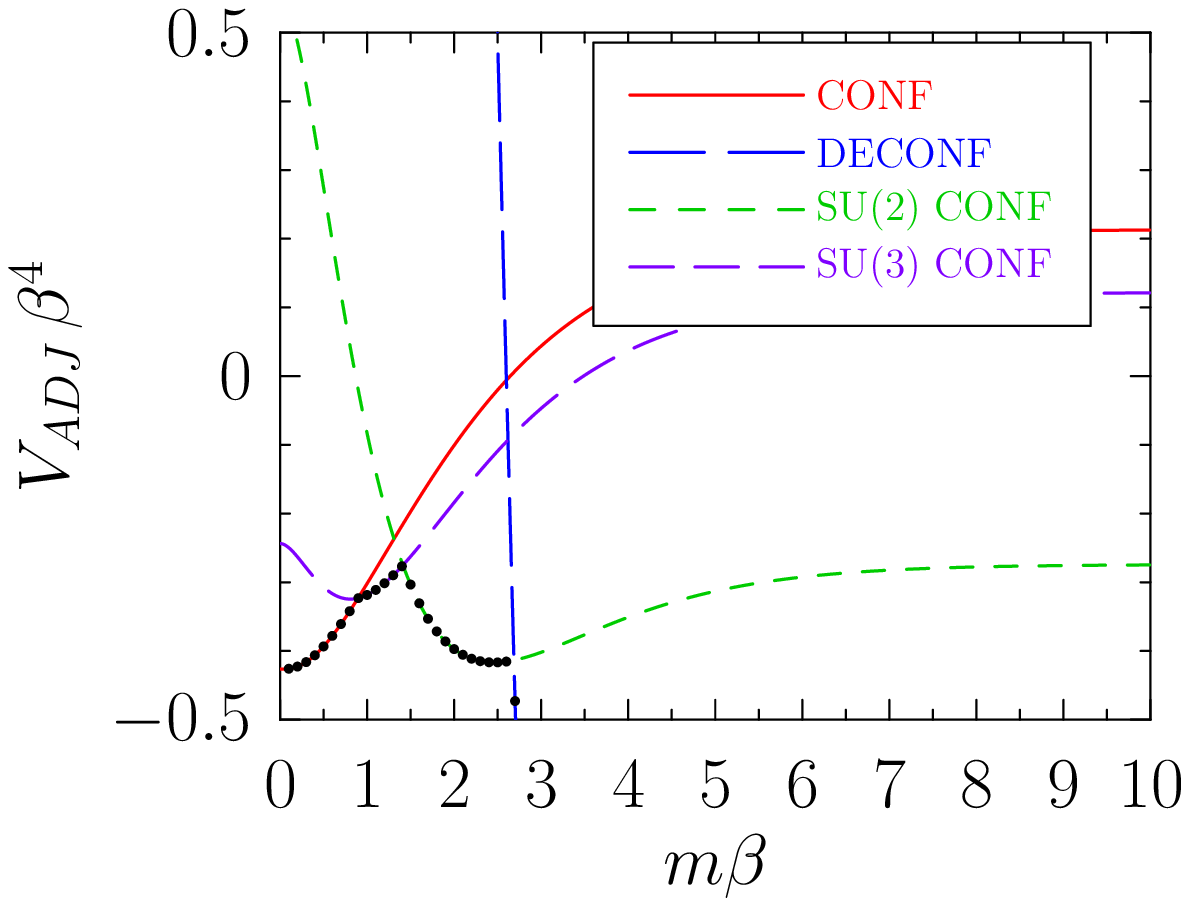}
%      \caption{$V_{ADJ(+)}$ for $N = 6$, $N_f = 3$ Majorana flavours}
%      \label{adj_n6_nf3}
    \end{center}
  \end{minipage}
  \hfill
  \caption{$N = 6$: $V_{ADJ(+)}$ for (Left) $N_f = 2$; (Right) $N_f = 3$ Majorana flavours}
  \label{adj_n6}
\end{figure}

\begin{figure}
    \begin{center}
\includegraphics[width=6cm]{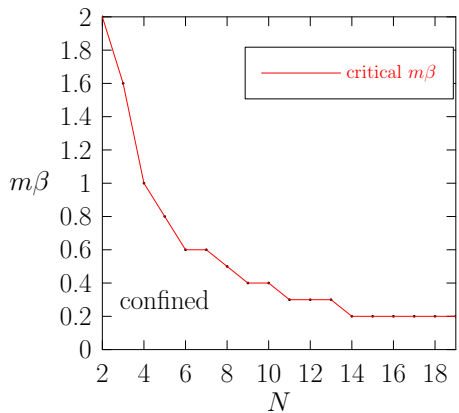}
       \caption{Range of $m \beta$ for which the confined phase is accessible in QCD(Adj) with $N_f = 2$}
       \label{conf}
    \end{center}
\end{figure}

\end{subsection}
%%%%%%%%%%%%%%%%%%%%%%%%%%%%%%%%%%%%%%%%%%%
\end{section}
\begin{section}{Conclusions}
%%%%%%%%%%%%%%%%%%%%%%%%%%%%%%%%%%%%%%%%%%
Extending Yang-Mills theory with adjoint fermions (with PBC) or using the center-stabilzing potential results in exotic phase structure. The center-stabilized theory and adjoint QCD with $N_f \ge 2$ Majorana flavours leads to perturbative access to the confined phase for all $N$. For adjoint QCD with at least two Majorana fermion flavours, as $N$ increases the range of $m \beta$ for which the confined phase is accessible decreases. However, as $N_f$ is increased within the limits allowed by asymptotic freedom, the confined phase becomes accessible for a larger range of $m \beta$. The center-stabilized theory contains all the phases of adjoint QCD and these can be traversed in the $a_n$ parameter space in the same order as they appear when increasing $m \beta$ in adjoint QCD, avoiding extraneous phases. Considering also the results in \cite{Myers:2008jm}: for $QCD (AS/S)$ with PBC on fermions a ${\cal C}$-breaking phase is favoured for all $m \beta$.
%%%%%%%%%%%%%%%%%%%%%%%%%%%%%%%%%%%%%%%%%%%
\end{section}

\end{document}